\documentclass[aps,prl,print,twocolumn,showpacs,longbibliography]{revtex4-1} 
\pdfoutput=1
\usepackage{amsthm}
\usepackage{amsmath}
\usepackage{amsfonts,amssymb}
\usepackage{graphicx}

\usepackage{verbatim}

\usepackage{mathtools} 

\usepackage{epigraph}

\usepackage{tabularx}
\newcolumntype{D}{>{\centering\arraybackslash}X}

\usepackage{hyperref}
\hypersetup{
    bookmarksnumbered=true, 
    unicode=false, 
    pdfstartview={FitH}, 
    pdftitle={}, 
    pdfauthor={}, 
    pdfsubject={}, 
    pdfcreator={}, 
    pdfproducer={}, 
    pdfkeywords={}, 
    pdfnewwindow=true, 
    colorlinks=true, 
    linkcolor=blue, 
    citecolor=blue, 
    filecolor=blue, 
    urlcolor=blue 
}

\usepackage{epstopdf}

\usepackage{bbold} 

\newcommand{\ket}[1]{|#1\rangle}
\newcommand{\bra}[1]{\langle#1|}

\newtheorem{theorem}{Theorem}

\usepackage[usenames,dvipsnames]{color}
\usepackage{xcolor}

\def\undertilde#1{\mathord{\vtop{\ialign{##\crcr
$\hfil\displaystyle{#1}\hfil$\crcr\noalign{\kern1.5pt\nointerlineskip}
$\hfil\tilde{}\hfil$\crcr\noalign{\kern1.5pt}}}}}

\begin{document}

\title{Optimal Hamiltonian Simulation by Quantum Signal Processing}

\author{Guang Hao Low, Isaac L. Chuang}
\affiliation{Department of Physics, Center for Ultracold Atoms, and Research Laboratory of Electronics
\\
 Massachusetts Institute of Technology, Cambridge, Massachusetts 02139, USA}

\date{\today}

\begin{abstract}
The physics of quantum mechanics is the inspiration for, and underlies, quantum computation. As such, one expects physical intuition to be highly influential in the understanding and design of many quantum algorithms, particularly simulation of physical systems. Surprisingly, this has been challenging, with current Hamiltonian simulation algorithms remaining abstract and often the result of sophisticated but unintuitive constructions. We contend that physical intuition can lead to optimal simulation methods by showing that a focus on simple single-qubit rotations elegantly furnishes an optimal algorithm for Hamiltonian simulation, a universal problem that encapsulates all the power of quantum computation. Specifically, we show that the query complexity of implementing time evolution by a $d$-sparse Hamiltonian $\hat{H}$ for time-interval $t$ with error $\epsilon$ is $\mathcal{O}(td\|\hat{H}\|_{\text{max}}+\frac{\log{(1/\epsilon)}}{\log{\log{(1/\epsilon)}}})$, which matches lower bounds in all parameters. This connection is made through general three-step ``quantum signal processing'' methodology, comprised of (1) transducing eigenvalues of $\hat{H}$ into a single ancilla qubit, (2) transforming these eigenvalues through an optimal-length sequence of single-qubit rotations, and (3) projecting this ancilla with near unity success probability.
\end{abstract} 


\maketitle




\epigraph{
``If you want to make a simulation of nature, you'd
better make it quantum mechanical, and by golly it's a wonderful problem,
because it doesn't look so easy.''
}{\textit{Richard P. Feynman}~\cite{Feynman1982}}

\textbf{\textit{Introduction}} -- 
Quantum computers are based on the physics of quantum mechanics, a fundamental tenant of Nature as we know it. Thus it seems natural to expect that the design and interpretation of quantum algorithms be heavily driven by physical intuition. The adiabatic algorithm~\cite{farhi2000quantum,Mizel2007Adiabatic} inspired by adiabaticity, and quantum walks~\cite{Childs2009Walk,Childs2013UniversalWalk} inspired by locality, are prominent examples.  However, many quantum algorithms, most surprisingly those for the simulation of physical systems~\cite{Georgescu2014Simulation}, are not as similarly insightful, and successive improvements in their complexity and analysis trend towards increasing abstraction and mathematical sophistication.

Analogous to physical theories, good quantum algorithms for physics simulations should, beyond being correct, also ideally be simple. In seeking simplicity, not only is their implementation on physical machines eased, but so too could their performance and understanding be enhanced. As the essence of coherent quantum computation is the design of unitary operations with desired properties, this motivates  consideration of its closest analogue in experiments: physical quantum control~\cite{Khodjasteh2010}, which has a similar goal of designing quantum \textit{response functions}~\cite{Low2016methodology}. 

This hints at a deep connection between the design of optimal quantum algorithms and the synthesis of optimal quantum control policies. While robust time-optimal control~\cite{Caneva2009,Khodjasteh2009} is certainly an established tool in quantum computing, its role is often secondary to the ends: the synthesis of computing primitives, such as Clifford gates or even the quantum Fourier transform. It would be more desirable if physical dynamics were directly applicable to generic quantum algorithms without this intermediary. Indeed, the fact that intuition of the simplest quantum control -- discrete single-qubit rotations $\hat{R}_\phi(\theta)=e^{-i\frac{\theta}{2}(\hat{\sigma}_x\cos{\phi}+\hat{\sigma}_y\sin{\phi})}$ -- can extend to algorithms such as Grover search supports this notion. 

This relationship is made concrete by interpreting discrete sequences of physical operations as  \textit{programs} that compute functions. In the simplest setting, chaining $N$ identical rotations generates $\bra{1}\hat{R}_{\pi/2}^N(\theta)\ket{0}=\sin{(N\theta/2)}$. With $\theta$ as the input, this computes the function $f(\theta)=\sin{(N\theta/2)}$, which may be estimated through measurement. As Pauli matrices 
 $\hat{\sigma}_{x,y,z}$ form a complete basis for $2$-by-$2$ matrices, generic sequences of the form
\begin{align}
\label{eq:ResponseFunction}
\hat{V}(\theta)&=\hat{R}_{\phi_N}(\theta)\hat{R}_{\phi_{N-1}}(\theta)\cdots \hat{R}_{\phi_1}(\theta), \quad \vec{\phi}\in \mathbb{R}^N, \\ \nonumber
&=A(\theta)\hat{\mathbb 1}+iB(\theta)\hat{\sigma}_z+iC(\theta)\hat{\sigma}_x+iD(\theta)\hat{\sigma}_y,
\end{align}
which we fully characterized in~\cite{Low2016methodology}, then compute more general functions of $\theta$ in the real $A,B,C,D$, through a program specified by some choice of phases $\vec{\phi}$. In fact, the isomorphism of single-qubit rotations to those on a sphere (up to a double covering), furnishes an intuitive \textit{classical} interpretation for this simple model of quantum computation. Moreover, the quantum control in Eq.~\ref{eq:ResponseFunction}, being piecewise, is naturally compatible with the inherently discrete nature of fault-tolerant architectures.

Though physically appealing, the computational merit of directly exploiting the structure of single-qubit rotations, or any physical system, ultimately rests on two criteria: \textbf{(1)} usefulness in solving important problems, and \textbf{(2)} optimality in space and time resources. It is also particularly challenging to see how this approach could apply generally to the complex multi-qubit dynamics arising in the simulation of quantum systems.

The simulation problem maps one set of physical dynamics of interest -- described by Hamiltonian $\hat{H}$ -- to another physical system that can be precisely controlled. Thus one expects  the role of physics to be preeminent and obvious. Following seminal work by Lloyd~\cite{Seth1996universal} for Hamiltonians with local interactions, and Aharonov and Ta-Shma~\cite{Aharonov2003Adiabatic} for more general sparse Hamiltonians, many celebrated results have been obtained over the years~\cite{Berry2006,Childs2010,Childs2011,Childs2012,Berry2014,Berry2015Hamiltonian,Berry2015Truncated} for approximating the time evolution operator $e^{-i\hat{H}t}$ for time-interval $t$ with error $\epsilon$. Encouragingly, intuitive quantum walks are already part of state-of-art. More, however, could be hoped from their other components.

The complexity of such quantum algorithms is usually judged by the number of queries made to a unitary quantum oracle $\hat{O}$ that provides a description of $\hat{H}$. Many interesting physical system are described by the especially well-studied model of $d$-sparse $\hat{H}$ with at most $d$ non-zero elements in every row, and the best known algorithms~\cite{Berry2015Hamiltonian} are based on the Childs quantum walk~\cite{Childs2010}, which builds upon the Szegedy walk~\cite{Szegedy2004spectra}, that simulates time evolution by $\arcsin{(\hat{H})}$ which must be linearized. The difficulty lies in finding a quantum circuit that does this with the fewest queries to $\hat{O}$ and the fewest number of additional primitive quantum gates.

Lower bounds on the query cost are well-known. The ``no-fast-forwarding'' theorem~\cite{Berry2006,Berry2015Hamiltonian} demands at least $\Omega(\tau)$ queries independent of $\epsilon$, where $\tau=td\|\hat{H}\|_{\text{max}}$ and $\|\hat{H}\|_{\text{max}}$ is the largest element of $\hat{H}$ in absolute value, and impressive recent work~\cite{Berry2014,Berry2015Hamiltonian} proved an exact error scaling of $\Theta\big(\frac{\log{(1/\epsilon)}}{\log{\log{(1/\epsilon)}}}\big)$ for $\tau=\mathcal{O}(1)$. Though this suggests a naive additive lower bound $\Omega\big(\tau+\frac{\log{(1/\epsilon)}}{\log{\log{(1/\epsilon)}}}\big)$~\cite{Berry2015Hamiltonian}, the best algorithms to date approach these factors multiplicatively with either linear scaling in time $\mathcal{O}(\frac{\tau}{\sqrt{\epsilon}})$~\cite{Childs2010} or sub-logarithmic scaling in error $\mathcal{O}\big(\tau\frac{\log{(\tau/\epsilon)}}{\log{\log{(\tau/\epsilon)}}}\big)$~\cite{Cleve2009,Berry2014}. Long unanswered is the existence of an algorithm that is \emph{additively} optimal, with implications for the relation between continuous and discrete-time models of physics, and of interest in problems~\cite{Sussman1988} where $\tau,\epsilon$ scale together.

We achieve precisely this with a simple algorithm that matches the additive lower bound. In fact, it also realizes the optimal trade-off between time and error, thus no further improvement in query complexity for this formulation of Hamiltonian simulation is possible. 
Compared to prior art~\cite{Childs2010,Berry2014}, this represents up to a square-root improvement. Moreover, the space overhead in ancilla qubits, beyond those required for the quantum walk, is reduced from scaling with some function of $\tau/\epsilon$ to just $1$. 

\begin{figure}
\includegraphics[width=1.0\columnwidth]{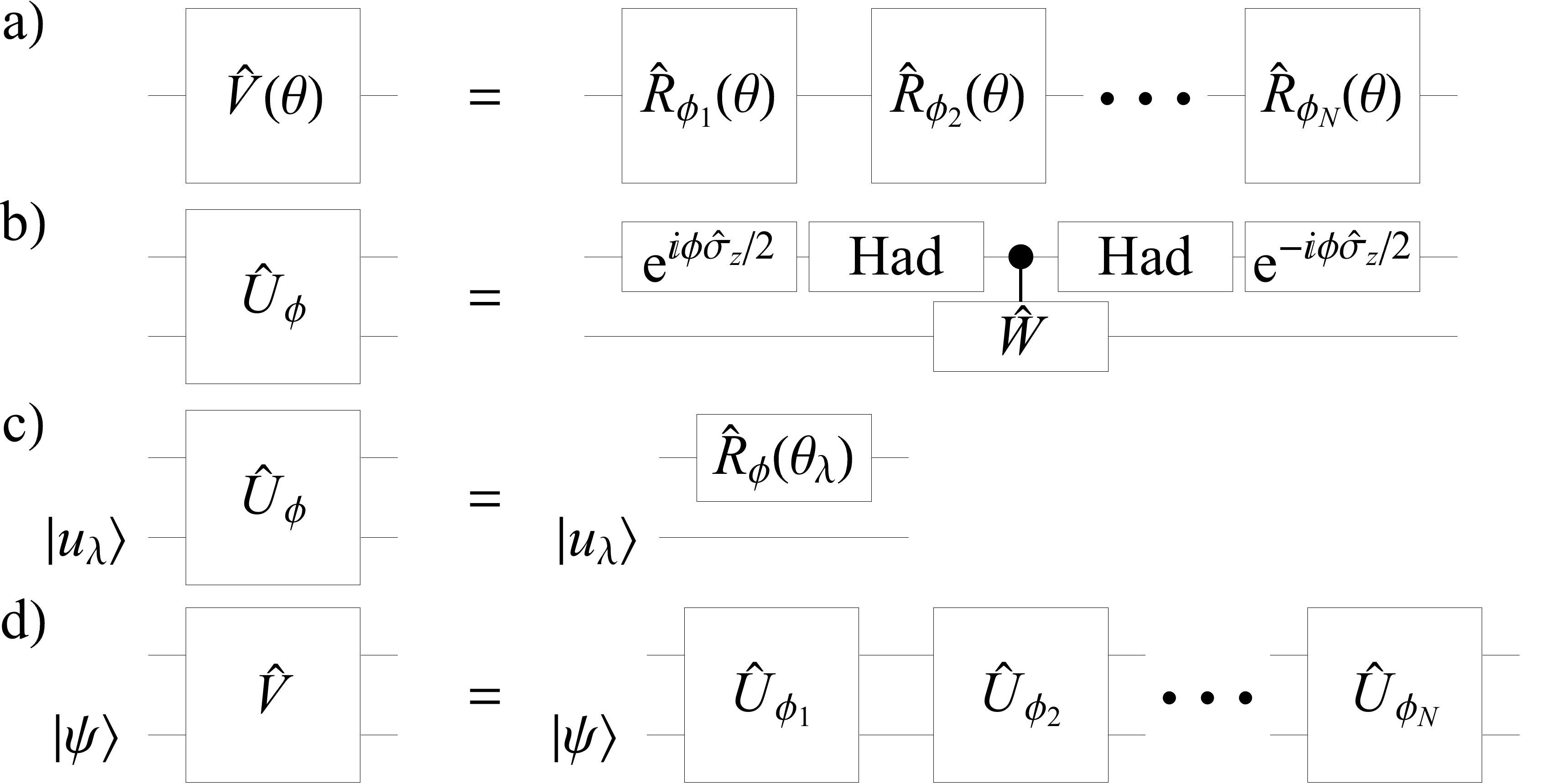}
\caption{\label{fig:circuits}Quantum circuits mapping (a) a sequence of single-qubit rotations $\hat{V}(\theta)$ to (d) quantum signal processing $\hat{V}$. Each single-qubit rotation $\hat{R}_\phi(\theta)$ is replaced by (b) $\hat{U}_\phi$, built from Hadamard gates and controlled-$W$ with eigenstates $\hat{W}\ket{u_\lambda}=e^{i\theta_\lambda}\ket{u_\lambda}$. Thus (c) $\hat{U}_\phi$ on input $\ket{u_\lambda}$ reduces to a single-qubit rotation $\hat{R}_\phi(\theta_\lambda)$. By linearity, $\hat{V}$ on an arbitrary input $\ket{\psi}$ may be understood as rotations $\hat{V}(\theta_\lambda)$ controlled by a superposition of $\ket{u_\lambda}$. By some choice of single-qubit input state and measurement basis, coefficients of the $\ket{u_\lambda}$ are then rescaled by the components of the function $\hat{V}(\theta_\lambda)$ programmed by $\vec{\phi}$.}
\end{figure}

Most remarkably, this is achieved by finding a class of computational problems addressed by the optimal control of the single-qubit in Eq.~\ref{eq:ResponseFunction} in a very natural way. Given a unitary $\hat{W}$ with eigenstates $\hat{W}\ket{u_\lambda}=e^{i\theta_\lambda}\ket{u_\lambda}$, we consider the general problem of constructing a quantum circuit $\hat{V}_{\text{ideal}}$ with transformed eigenphases
\begin{align}
\label{eq:HamSimTargetfunction}
\hat{W}
\mapsto
\hat{V}_{\text{ideal}}=\sum_{\lambda}e^{i h(\theta_{\lambda})}\ket{u_\lambda}\bra{{u_\lambda}},
\end{align}
using the fewest queries to controlled-$\hat{W}$ for any real function $h(\theta)$. We call our solution to this ``quantum signal processing'' (Fig.~\ref{fig:circuits}), and its application to $d$-sparse Hamiltonian simulation leads to tremendous simplification and the claimed improvements. Our success here elevates optimal discrete quantum control in general as a tool that can be rigorous and essential in the design of optimal quantum algorithms, thus providing a medium through which physical intuition may flow. 
 
Two key properties distinguish quantum signal processing from routines that can effect similar transformations, such as quantum phase estimation~\cite{Nielsen2004} or linear-combination-of-unitaries~\cite{Childs2012,Berry2015Hamiltonian,Berry2015Truncated} which require a large number of ancilla. First, is its intuitive use of just a single ancilla qubit. Second, the query complexity of the methodology is exactly the degree $N$ of optimal trigonometric polynomial approximations to $e^{i h(\theta)}$ with error $\epsilon$~\cite{Mcclellan1973,Pachon2009,powell1981approximation,trefethen2013approximation}, without the decaying success probability of prior art. Analogous to digital filter design techniques in discrete-time signal processing~\cite{oppenheim2010} this also elegantly bridges the design of a number of quantum algorithms to the vast field of function approximation~\cite{powell1981approximation}.

In the following, we describe the reduction of quantum signal processing to optimal quantum control, and show how to efficiently choose the phases $\vec{\phi}$ Eq.~\ref{eq:ResponseFunction} such that any unitary transformation Eq.~\ref{eq:HamSimTargetfunction} is approximated with error $\epsilon$ and success probability $1-\mathcal{O}(\epsilon)$. The essential features of the quantum walk are then reviewed to show how quantum signal processing for the special case of $h(\theta)=-\tau \sin{(\theta)}$ solves the sparse Hamiltonian simulation problem. 
That this achieves lower bounds follows by analyzing the scaling between $N,\tau,\epsilon$ of quantum signal processing for this $h(\theta)$.

\textbf{\textit{Quantum signal processing}} -- All quantum algorithms require a rigorous analysis of their resource costs in space and time. Thus any form of quantum control repurposed to such ends must have a similarly rigorous characterization. Previously~\cite{Low2016methodology}, we studied the optimal control of arbitrary sequences of single-qubit rotations in Eq.~\ref{eq:ResponseFunction}, provided an intuitive characterization of the functions $A,B,C,D$ achievable by some choice of $\vec{\phi}$, and provided efficient algorithms for synthesizing all these functions and the required $\vec{\phi}$ from some partial specification. The results relevant here are:
\begin{theorem} [Achievable ($A$,$C$)] 
\label{thm:AchievablePartialTuples}
$\forall$ even $N>0$, a choice of real functions $A(\theta),C(\theta)$ can be implemented by some $\vec{\phi}\in\mathbb{R}^N$ if and only if all these are true:\\
(1) $\forall \theta\in\mathbb{R},\;A^2(\theta)+C^2(\theta)\le 1$.$\quad$ (2) $A(0)=1$.\\
(3) $A(\theta)=\sum^{N/2}_{k=0}a_k \cos{(k\theta)}, \quad\{a_k\}\in \mathbb{R}^{N/2+1}$.\\
(4) $C(\theta)=\sum^{N/2}_{k=1}c_k \sin{(k\theta)}, \quad\{c_k\}\in \mathbb{R}^{N/2}$.\\
Moreover, $\vec{\phi}$ can be efficiently computed from $A(\theta),C(\theta)$.
\end{theorem}
Note that in~\cite{Low2016methodology}, $A, C$ are expressed as trigonometric polynomials, but can be rewritten a Fourier series (3), (4) using Chebyshev polynomials of the first and second kind $T_{k}(\cos{(\theta)})=\cos{(k\theta)}$ and $U_{k}(\cos{\theta})=\frac{\sin{((k+1)\theta)}}{\sin{\theta}}$. 

We now map these results, in three steps, to quantum signal processing which transforms an arbitrary input unitary $\hat{W}=\sum_\lambda e^{i\theta_{\lambda}}\ket{u_{\lambda}}\bra{u_{\lambda}}$ into one with modified eigenphases $\hat{V}_{\text{ideal}}=\sum_{\lambda}e^{i h(\theta_{\lambda})}\ket{u_\lambda}\bra{{u_\lambda}}$: 

\textbf{(a) Signal transduction} of $\hat{W}$ into a signal unitary classically controlled by $\phi\in\mathbb{R}$:
\begin{align}
\label{eq:Uoracle}
\textstyle \hat{U}_{\phi}=\sum_{\lambda}\hat{R}_{\phi}(\theta_{\lambda})\otimes\ket{{u_\lambda}}\bra{{u_\lambda}}.
\end{align}
This is implemented in Fig.~\ref{fig:circuits}b with one controlled-$\hat{W}$, which is always possible on a quantum computer in the worst-case by replacing all of its gates with controlled version, and $\mathcal{O}(1)$ single-qubit rotations:
\begin{align}
\label{eq:OracleQuerySynthesis}
\hat{U}_\phi&=(e^{-i\phi\hat{\sigma}_z/2}\otimes \hat{\mathbb{1}})\hat{U}_0 (e^{i\phi\hat{\sigma}_z/2}\otimes \hat{\mathbb{1}}), \\ \nonumber
\hat{U}_0&=\ket{+}\bra{+}\otimes\hat{\mathbb 1}+\ket{-}\bra{-}\otimes\hat{W} \\ \nonumber
&\textstyle=\sum_{\lambda}e^{i\theta_{\lambda}/2}\hat{R}_0(\theta_{\lambda})\otimes\ket{u_\lambda}\bra{u_{\lambda}},
\end{align} 
where $\ket{\pm}=\frac{\ket{0}+\ket{1}}{\sqrt{2}}$. As $\hat{U}_{\phi}$ acting on $\ket{u_\lambda}$ selects the rotation $\hat{R}_{\phi}(\theta_{\lambda})=\bra{u_\lambda}\hat{U}_{\phi}\ket{u_\lambda}$ as seen in Fig.~\ref{fig:circuits}c, these are precisely the single-qubit ancilla rotations in Eq.~\ref{eq:ResponseFunction} with rotation angle $\theta_\lambda$ controlled by the $\lambda$ index, but with an additional global phase $e^{i\theta_{\lambda}/2}$.

\textbf{(b) Signal transformation} by computing unitary functions $\hat{V}(\theta_\lambda)$ over a superposition of $\theta_{\lambda}$ on the single-qubit ancilla through the simple circuit of Fig.~\ref{fig:circuits}d:
\begin{align}
\label{eq:QSPquantumcircuit}
\hat{V}&=\hat{U}_{\phi_N}\hat{U}_{\phi_{N-1}}\cdots\hat{U}_{\phi_1},\quad \vec{\phi}\in\mathbb{R}^{N}, \; N\text{ even}.
\end{align}
As this invokes $\hat{W}$ a number $N$ times, its query cost is $\mathcal{O}(N)$. Note that the unwanted phase $e^{i\theta_{\lambda}/2}$ can be uncomputed by alternating between $\hat{U}_{\phi}$ and $\hat{U}^{\dag}_{\phi+\pi}$
since $\hat{R}_\phi(\theta)=\hat{R}^{\dag}_{\phi+\pi}(\theta)$ and $N$ is even. 

\textbf{(c) Signal projection} of the ancilla onto some basis, to select desired components of $\hat{V}(\theta_\lambda)$ in Eq.~\ref{eq:ResponseFunction}. As the desired phase transformation can be implemented through $A(\theta),C(\theta)$, Consider the input state $\ket{+}\ket{u_\lambda}$, and postselect on measuring $\bra{+}$. Other choices are of course possible. This applies onto state $\ket{u_\lambda}$ the coefficient
\begin{align}
\label{eq:PhaseShift}
\bra{+}\hat{V}\ket{+}\ket{{u_\lambda}}&=
\left(A(\theta_{\lambda})+i C(\theta_{\lambda})\right)\ket{u_{\lambda}},\\
p&=
\min_{\theta\in\mathbb{R}}|\bra{+}\hat{V}(\theta)\ket{+}|^2=\min_{\theta\in\mathbb{R}}|A(\theta)+i C(\theta)|^2, \nonumber
\end{align}
with worst-case success probabilitiy $p$. Thus \textbf{(a)}-\textbf{(c)} provide a reduction from finding quantum algorithms for approximating $\hat{V}_{\text{ideal}}$ to finding Fourier approximations of $A(\theta)+i C(\theta)$ to $e^{i h(\theta)}$. 

By applying Thm.~\ref{thm:AchievablePartialTuples} to these three steps of quantum signal processing, we now prove following theorem which furnishes the complexity of implementing $\hat{V}_{\text{ideal}}$ given this Fourier approximation:
\label{sec:QSP}
\begin{theorem}[Quantum Signal Processing]
\label{thm:QSP}
$\forall$ real odd periodic functions $h:(-\pi,\pi]\rightarrow (-\pi,\pi]$ and even $N>0$, let $(A(\theta),C(\theta))$ be real Fourier series in $(\cos{(k\theta)},\sin{(k\theta)}),\;k=0,...,N/2$, that approximate
\begin{align}
\label{eq:FourierError}
\max_{\theta\in\mathbb{R}}|A(\theta)+i C(\theta)-e^{i h(\theta)}|\le \epsilon.
\end{align}
Given $A(\theta),C(\theta)$, one can efficiently compute the $\vec{\phi}$ such that $\bra{+}\hat{V}\ket{+}$ in Eq.~\ref{eq:QSPquantumcircuit} applies $\hat{U}_{\phi}$ a number $N$ times to approximate $\hat{V}_{\text{ideal}}$ in Eq.~\ref{eq:HamSimTargetfunction} with success probability $p\ge 1-16 \epsilon$ and trace distance 
\begin{align}
\label{eq:VError}
\epsilon_{\text{Tr}}=\max_{\ket{\psi}}\|(\bra{+}\hat{V}\ket{+}-\hat{V}_{\text{ideal}})\ket{\psi}\|\le 8 \epsilon.
\end{align}
\end{theorem}
Note that the restricted symmetry of $h(\theta)$ is for consistency with the parity and periodicity of $A(\theta),C(\theta)$.

Given $A(\theta),C(\theta)$ that satisfy Eq.~\ref{eq:FourierError}, conditions (1), (2) of Thm.~\ref{thm:AchievablePartialTuples} will not generally be satisfied. (1) is violated as $\max_{\theta}(|A(\theta)|,|C(\theta)|)\le 1+\epsilon$. Thus we 
rescale
\begin{align}
\label{eq:A1function}
&A_1(\theta)=A(\theta)/(1+\epsilon),\quad C_1(\theta)=C(\theta)/(1+\epsilon), \\\nonumber
&|(A_1(\theta)+iC_1(\theta))-e^{i h(\theta)}|\le \epsilon/(1+\epsilon)+\epsilon\le 2\epsilon,
\end{align}
at the cost of a slightly larger error $2\epsilon$.
Note that $A^2_1+C^2_1\ge\big(\frac{1-\epsilon}{1+\epsilon}\big)^2$.
(2) is violated as $A_1(0)=\cos{\delta}\ge \frac{1-\epsilon}{1+\epsilon}$ for some $\delta\in\mathbb{R}$. Fixing this is more involved. As $\hat{V}(\theta)$ is unitary, $A_1^2+B^2+C_1^2+D^2=1$. We can apply the prescription in~\cite{Low2016methodology} using polynomial sum-of-squares to compute the unspecified $B,D$ from $A_1,C_1$ such that $B,D$ are of the form (3) and (4) respectively. Thus $A_1^2(0)+B^2(0)=1$, and $|B(0)|=|\sin{\delta}|$. Define 
\begin{align}
\label{eq:A2function}
A_2(\theta)&=A_1(\theta)\cos{\delta}+B(\theta)\sin{\delta}, \\ \nonumber
|A_2(\theta)-A_1(\theta)|&\le\frac{2\epsilon}{1+\epsilon}+|B(\theta)|\frac{2\sqrt{\epsilon}}{1+\epsilon}\le 6\epsilon.
\end{align}
This introduces an additional error by using the triangle inequality and $B^2\le 1-A_1^2-C_1^2\le 1-\big(\frac{1-\epsilon}{1+\epsilon}\big)^2 = \frac{4\epsilon}{(1+\epsilon)^2}$. By construction, $A_2(0)=1$. The functions $A_2(\theta),C_1(\theta)$ thus satisfy Thm.~\ref{thm:AchievablePartialTuples}. By adding the errors in  Eqs.~\ref{eq:A1function},~\ref{eq:A2function},
the distance of $\bra{+}\hat{V}\ket{+}$ from $\hat{V}_{\text{ideal}}$ in Eq.~\ref{eq:HamSimTargetfunction} and the worst-case success probability in Eq.~\ref{eq:PhaseShift} are
\begin{align}
\label{eq:VTraceDistanceEror}
\epsilon_{\text{Tr}}&\le\max_{\theta\in\mathbb{R}}|A_2(\theta)+i C_1(\theta)-e^{i h(\theta)}|\le 8\epsilon, \\ \nonumber
p&\ge (1-8\epsilon)^2\ge 1-16\epsilon.
\end{align}

The optimality of this remarkably simple procedure for eigenphase transformation follows from its role in obtaining the best possible quantum algorithm for $d$-sparse Hamiltonian simulation. We now highlight essential features of a quantum walk $\hat{W}$ constructed from the oracles $\hat{O}$ describing sparse $\hat{H}$.

\textit{\textbf{Childs' quantum walk}}~\cite{Childs2010} -- 
$\hat{W}$ can be constructed from oracles that specify a $d$-sparse Hamiltonian $\hat{H}$ with  $n$-qubit eigenstates $\hat{H}\ket{\lambda}=\lambda\ket{\lambda}$.
Access to two oracles $\hat{O}_{H},\hat{O}_{F}$ is commonly assumed: 
$\hat{O}_{H}$ accepts the input $(j,k)\in[2^n]\times [2^n]$ on $2n$-qubit registers and returns $H_{j,k}=\bra{j}\hat{H}\ket{k}$ in another $m$-qubit register.
$\hat{O}_{F}$ accepts the input $(j,l)\in[2^n]\times [d]$ on the same $2n$-qubit registers and computes in place the column index $f(j,l)\in[2^n]$ of the $l^{\text{th}}$ nonzero element in the $j^{\text{th}}$ row of $\hat{H}$. 

It is well-known~\cite{Berry2012} that with $1$ query to $\hat{O}_F, \hat{O}_H,\hat{O}^{\dag}_H$ each and $\mathcal{O}(n+m \text{ poly}(\log m))$ primitive gates, one can implement an isometry $\hat{T}$ that maps every state $\ket{\lambda}\ket{0}^{\otimes n+m+2}$ onto two eigenstates $\ket{\lambda\pm}$ of $\hat{W}$:
\begin{align}
\hat{T}\ket{\lambda}&=\left(\ket{\lambda +}+\ket{\lambda-}\right)/\sqrt{2}.
\end{align}
Moreover, $\hat{T}$ is constructed such that the walk $\hat{W}=i\hat{S}(2\hat{T}\hat{T}^{\dag}-\hat{\mathbb 1})$ has eigenvalues $\hat{W}\ket{\lambda\pm}=e^{i\theta_{\lambda\pm}}\ket{\lambda\pm}$,
\begin{align}
\theta_{\lambda{\pm}}&=\pm \arcsin{(\lambda/\|\hat{H}\|_{\text{max}} d)}+(1\mp 1)\pi/2,
\end{align}
that depends on the $\hat{H}$ eigenvalues $\lambda$. As $\hat{W}$ corresponds to reflection about $\hat{T}\hat{T}^{\dag}$ followed by swapping $(2n+2)$-qubit registers with $\hat{S}$, its query and gate complexities are identical to $\hat{T}$ up to constant factors.

Hamiltonian simulation is achieved by creatively applying $\hat{W}$ some number of times to implement $\ket{\lambda\pm}\mapsto e^{-i\lambda t}\ket{\lambda\pm}$, independent of the $\pm$ index. Uncomputing with $\hat{T}^{\dag}$ then maps $\ket{\lambda\pm}$ back onto $\ket{\lambda}\ket{0}^{\otimes n+m+2}$ with the desired phase evolution. However, some difficulties arise. First, the applied phase $\theta_{\lambda\pm}$ is nonlinear in $\lambda$. Second, each eigenstate $\ket{{\lambda\pm}}$ evolves under $\hat{W}$ with phases in opposite directions. Thus uncomputing with $\hat{T}^{\dag}$ does not map $\hat{W}\hat{T}\ket{\lambda}\ket{0}^{\otimes n+m+2}$ back onto the basis $\ket{\lambda}\ket{0}^{\otimes n+m+2}$. 
In~\cite{Berry2015Hamiltonian}, these are overcome by approximating the unitary transformation in Eq.~\ref{eq:HamSimTargetfunction} with target function
\begin{align}
\label{eq:Htargetfunction}
h(\theta)=-\tau\sin{(\theta)}\; \Rightarrow \; h(\theta_{\lambda\pm})=- \lambda t,
\end{align}
resulting in the desired phase, but implemented using a technique combining a linear combination of $N$-controlled $\hat{W}^{1,...,N}$ such that the success probability decays with $N$. Our quantum signal processing methodology, does not experience such a decay and its direct application furnishes an optimal Hamiltonian simulation algorithm

\textbf{\textit{Hamiltonian Simulation}} --
\label{sec:HamSim}
Applying quantum signal processing in Thm.~\ref{thm:QSP} to Hamiltonian simulation requires a good Fourier approximation to
\begin{align}
A(\theta)+iC(\theta)\approx e^{i h(\theta)}=e^{-i \tau\sin{(\theta)}},
\end{align}
which is provided by the Jacobi-Anger expansion~\cite{Abramowitz1966handbook}
\begin{align}
\label{eq:Jacobi-Anger}\nonumber
\cos{( \tau \sin{(\theta)})}&=\textstyle J_{0}(\tau)+2\sum^{\infty}_{k\text{ even}>0}J_{k}(\tau)\cos{(k \theta)},\\
\sin{\left( \tau \sin{(\theta)}\right)}&=\textstyle2\sum^{\infty}_{k\text{ odd}>0}J_{k}(\tau)\sin{(k \theta)}, 
\end{align}
where $J_{k}(\tau)$ are Bessel functions of the first kind. Note that these Fourier series are already in the form required by conditions (3), (4) of Thm.~\ref{thm:AchievablePartialTuples}. As $|J_{k}(\tau)|\le\frac{1}{|k|!}\left|\frac{\tau}{2}\right|^{|k|}$~\cite{Abramowitz1966handbook} decays rapidly with $k$, good approximations are obtained truncating Eq.~\ref{eq:Jacobi-Anger} at $k > N/2$. This approximates $e^{-i \tau\sin{(\theta)}}$ with error shown in~\cite{Berry2015Hamiltonian} for $\tau\le N/2=q-1$ to be
\begin{align}
\label{eq:TruncationError}
\epsilon \le \sum^{\infty}_{k=q}\displaystyle 2|J_{k}(\tau)|\le \frac{4\tau^{q}}{2^q q!}=\mathcal{O}\Big(\Big(\frac{e\tau}{2q}\Big)^q\Big).
\end{align}
Inserting into Thm.~\ref{thm:QSP}, the query complexity of Hamiltonian simulation follows by solving Eq.~\ref{eq:TruncationError} for $N$, using the implementation of $\hat{U}_{\phi}$ in Eq.~\ref{eq:OracleQuerySynthesis} with $\mathcal{O}(1)$ queries, and that $\hat{V}$ in Eq.~\ref{eq:QSPquantumcircuit} contains $N$ applications of $\hat{U}_{\phi}$. 

The optimality of this result for all input parameters follows from known lower bounds. Specifically, Eq.~\ref{eq:TruncationError} is matched with a corresponding lower bound $N=\Omega(q)$~\cite{Berry2015Hamiltonian,Kothari2016} for any $q$ satisfying
\begin{align}
\label{eq:ExactLowerBound}
\epsilon < \frac{1}{2}\Big|\sin{\Big(\frac{\tau}{q}\Big)}\Big|^q=\mathcal{O}\Big(\Big(\frac{\tau}{q}\Big)^q\Big).
\end{align}

Note that Eqs.~\ref{eq:TruncationError},~\ref{eq:ExactLowerBound} are solved by the Lambert $W$-function~\cite{Corless1996} which captures the detailed trade-off between $\tau$ and $\epsilon$. Its asymptotic behavior may be understood by substituting $q=\frac{e}{2}(\tau+\gamma)$, where $\tau,\gamma\ge 0$. When $\tau =\mathcal{O}(\gamma)$, one finds $\gamma=\mathcal{O}\big(\frac{\log{(1/\epsilon)}}{\log\log{(1/\epsilon)}}\big)$. Thus we express the complexity of Hamiltonian simulation as
\begin{theorem}[Optimal sparse Hamiltonian simulation]
\label{thm:QueryComplexity}
A $d$-sparse Hamiltonian $\hat{H}$ on $n$ qubits with matrix elements specified to $m$ bits of precision can be simulated for time-interval $t$, error $\epsilon$, and success probability at least $1-2\epsilon$ with $\mathcal{O}\big(td\|\hat{H}\|_{\text{max}}+\frac{\log{(1/\epsilon)}}{\log{\log{(1/\epsilon)}}}\big)$ queries and a factor $\mathcal{O}((n+m\text{polylog}(m)))$ additional quantum gates.
\end{theorem}
This is valid for $\tau=\mathcal{O}(\frac{\log{(1/\epsilon)}}{\log{\log{(1/\epsilon)}}})$ and stronger than prior art~\cite{Berry2014,Berry2015Hamiltonian} which assumes $\tau=\mathcal{O}(1)$. Unlike most Hamiltonian simulation algorithms, the query cost is additive in the simulation length $\tau$ and the target error $\epsilon$. As such, the $\tau$ term matches the lower bound $\Omega(\tau)$~\cite{Berry2006,Berry2015Hamiltonian} with no multiplicative dependence on error. 

\textit{\textbf{Conclusion}} -- 
\label{sec:Conclusion}
We have shown that optimal quantum algorithms for Hamiltonian simulation can be remarkably simple and physically-motivated. Here, physical intuition flows into the process by directly using the dynamics of discrete single-qubit rotations as a computational module, which proves to be exceptionally useful when translated into quantum signal processing. Indeed, we have focused on choosing target functions $(A,C)$ for even $N$ in Thm.~\ref{thm:AchievablePartialTuples}, but many other choices described in~\cite{Low2016methodology} are possible. For example, fixed-point amplitude amplification~\cite{Yoder2014} and Heisenberg-limited quantum imaging~\cite{Low2015} are special cases for the choice $(A(\theta)\propto T_{N}(\beta\cos(\theta/2)),B(\theta)=0),\beta>1$. 

Directly exploiting the structured dynamics in other physical systems could lead to powerful tools for a similarly intuitive approach to rigorous and optimal quantum algorithms. The question of what other important quantum algorithms can be designed or improved in this manner is an exciting natural extension to this work.

\textbf{\textit{Acknowledgements}} -- 
G.H. Low and I.L. Chuang thank Cedric Yen-Yu Lin, Robin Kothari, and Matthew Hastings for insightful discussions, and acknowledge funding by the ARO Quantum Algorithms Program, the NSF CUA, and NSF RQCC Project No.1111337.

\bibliography{PerfectHamiltonianSimulation}

\end{document}